\documentclass[journal]{IEEEtran}
\usepackage[dvips]{epsfig}
\usepackage[latin1]{inputenc}
\usepackage[T1]{fontenc}
\usepackage{booktabs}
\usepackage{amsmath}
\usepackage{amssymb}
\usepackage{array}

\def\BibTeX{{\rm B\kern-.05em{\sc i\kern-.025em b}\kern-.08em
    T\kern-.1667em\lower.7ex\hbox{E}\kern-.125emX}}

\begin{document}

\title{Multicasting in Cognitive Radio Networks: \\ \emph{Algorithms, Techniques and Protocols}}
\author{Junaid Qadir, Adeel Baig, Asad Ali, Quratulain Shafi
\thanks{This work has been supported by Higher Education Commission (HEC), Pakistan under the NRPU programme.}
\thanks{Junaid Qadir, Adeel Baig, Asad Ali, and Quratulain Shafi are with the Electrical Engineering Department at the School of Electrical Engineering and Computer Science (SEECS) at the National University of Sciences and Technology (NUST), Pakistan, \it{emails: junaid.qadir, adeel.baig, 12mseeaali, \newline10mscseqshafi@seecs.edu.pk}. }}
\maketitle

\begin{abstract}
Multicasting is a fundamental networking primitive utilized by numerous applications. This also holds true for cognitive radio networks (CRNs) which have been proposed as a solution to the problems that emanate from the static non-adaptive features of classical wireless networks. A prime application of CRNs is dynamic spectrum access (DSA), which improves the efficiency of spectrum allocation by allowing a secondary network, comprising of secondary users (SUs), to share spectrum licensed to a primary licensed networks comprising of primary users (PUs). Multicasting in CRNs is a challenging problem due to the dynamic nature of spectrum opportunities available to the SUs. Various approaches, including those based in optimization theory, network coding, algorithms, have been proposed for performing efficient multicast in CRNs. In this paper, we provide a self-contained tutorial on algorithms and techniques useful for solving the multicast problem, and then provide a comprehensive survey of protocols that have been proposed for multicasting in CRNs. We conclude this paper by identifying open research questions and future research directions.
\end{abstract}

\section{Introduction}

Cognitive radio networks (CRNs), networks of nodes equipped with cognitive radios (CRs), promise to revolutionize modern wireless networks by incorporating intelligence into its core \cite{akyildiz2009crahns}. CRs have themselves evolved from the concept of software-defined-radios (SDRs) that allowed a network to adapt to network conditions and user requirements in runtime with software changes only. CRs extend this concept to allow a node to observe its environment and to adapt to it through learning and cognition. A major use-case of CRNs is employing \emph{`dynamic spectrum access'} (DSA) to improve the wireless spectrum utilization. DSA has captured the fancy of industry, regulators, and academia as it promises to remedy the problem of inefficient spectrum utilization brought upon by the historical command-and-control approach to spectrum allocation. The main failing of this static spectrum allocation approach, which licensed specific portions of the radio spectrum to specific applications for exclusive usage, became apparent as more and more wireless technologies clamored for any available spectrum. It was observed that while most of the wireless spectrum was licensed, thereby being off-limits for newer innovative technologies that required spectrum as their lifeline, the spectrum was hardly utilized through a quirk of static spectrum policy and licensed user's bursty communication nature. DSA promises to solve this problem of `artificial spectrum scarcity' by allowing secondary users (SUs) access to the licensed spectrum subject to the condition that it does not cause any interference for the licensed, or the primary, users (PUs).

Multicasting or group communication is a fundamental networking primitive utilized by numerous wireless networking applications. Some envisioned applications include support of multimedia applications (such as video conferencing), file distribution, news or update dissemination, etc. \cite{paul2002survey} \cite{sahasrabuddhe2000multicast}. Multicasting subsumes the models of unicast transmission and broadcast transmission in its paradigm by varying the receiver group from the one extreme of a single receiver to the other extreme of all the network nodes as receivers.

Multicasting over wireless networks is a significant but challenging goal in the field of networks, which requires a lot of issues to be addressed like bandwidth, topology, loss of packets, routing, reliability, security issues and quality of service, before it can be deployed. Providing a trade-off between stability, throughput and packet loss with reduced bandwidth requirement and less power consumption is the main aim of multicast in wireless networks \cite{quinn2001ip}. Furthermore, multicasting is challenging in CRNs due to the dynamically changing topology of CRNs. CRNs often have to operate in unknown RF environments, with the topology of a secondary network depending critically on the temporal and spatial aspects of PU arrivals. This can lead to a scenario where the various CRN nodes have a heterogeneous set of channels available. This complicates the problem of multicast as the channel heterogeneity may mean lack of a common channel between neighbors \cite{akyildiz2006next}. More detailed analysis of routing challenges for multi-hop CRNs is provided in \cite{sengupta2013open}.

In this paper, we provide a tutorial on the general algorithms, techniques, and protocols that have been proposed for the wireless multicast problem, particularly, with a focus on CRNs. While there are numerous survey articles on wireless multicast \cite{varshney2002multicast}, optimization problems in multicast \cite{oliveira2005survey}, algorithm issues in multicast \cite{winter1987steiner}, this is the first work that coherently synthesizes necessary background from fields like optimization theory, network coding theory, algorithms, game-theory, machine-learning in the form of unified tutorial and follows it up with a survey of existing work on multicast in CRNs. We also provide a survey of the state-of-the-art in the field of multicast routing in CRNs, and highlight the open research challenges in this area.

\vspace{1mm}
\emph{Organization of the paper:} The rest of the paper is organized as follows. We provide the necessary background on multicasting, including defining basic terminology, various categories of multicasting approaches, and challenges of multicasting in section \ref{sec:background}. The challenges associated with multicast in CRNs are introduced in section \ref{sec:challengesmulticasting}. This is followed by a discussion on algorithms that have been proposed for addressing the wireless multicast problem in section \ref{sec:algorithm}. We then provide a self-contained tutorial on various techniques from fields like optimization, network coding, game-theory and machine learning in section \ref{sec:techniques}. A detailed survey of multicasting protocols proposed for CRNs is then provided in section \ref{sec:multicastprotocolsinCRNs}, followed by articulation of open research issues and future research directions in section \ref{sec:openissues}. Finally, this paper is concluded in section \ref{sec:conclusions}.

\section{Background: Multicasting}
\label{sec:background}

Before coming to our focused topic of algorithms, techniques and protocols for multicasting in CRNs, it is important to establish some background. In this section, we will introduce some basic multicasting terminology and concepts, multicasting goals, various challenges of multicasting, and categories of multicasting approaches in subsections \ref{sec:basicdefs}, \ref{sec:multicastinggoals}, \ref{sec:challengesmulticasting}, and \ref{sec:classificationmulticasting}, respectively.

\subsection{Multicasting Basics}
\label{sec:basicdefs}

The basic components of a multicast routing framework include sources, receivers, and groups having straightforward interpretations. The originator of the data stream is called a `source' while the end-user that wishes to receive the data stream is called a `receiver'. Receivers that share a common interest are pooled together in a multicast `group'. In multicast routing, the source is interested in transmitting its data stream to a group of hosts \cite{rosenberg2012primer}.

When there is a single receiver node, the routing problem is known as unicast routing which is typically solved by computing a shortest path between the source and the receiver. On the other extreme, the problem of transmitting the stream from the source to all the node in the network is known as broadcast routing or simply broadcasting.
Multicasting subsumes unicasting and broadcasting as a special case, and addresses the problem of reaching a group of nodes that collectively represent the destination.

To understand the various approaches that may be adopted to ensure packet delivery to a group of nodes, consider a simple example. Consider a multicasting example in which a source node $s$ wishes to send a stream of bandwidth $b$ to a group comprising $g$ other network nodes. A rather inefficient approach of doing this, known as \emph{unicast replication}, is to have $s$ create $g$ copies of the stream, i.e., a copy for each of the $g$ receivers, and to send each of these copies towards its destination node amongst the $g$ receivers using unicast. This approach is viable only when $b$ and $g$ are small and the intention is to avoid the overhead of implementing a multicast routing protocol. The disadvantage of using unicast replication is that there would be (unnecessarily)
increased consumption of bandwidth and general network resources.

For point-to-multipoint group communication, \emph{multicasting} comes across as the most efficient manner of communication. Different from replicated unicast, if a node is transmitting to multiple receivers through the same downstream neighbor, only one transmission is used in multicasting thereby suppressing the unnecessary redundancy and overhead. The \emph{efficiency} or gain of multicast in terms of network resource consumption compared to unicast has been studied in \cite{van2001efficiency}. Moreover, multicast also avoids the inefficiency of broadcasting when only a group of nodes
are intended receivers by sending data only to the interested nodes by creating multicast groups \cite{varshney2002multicast}.

\subsection{Multicasting Goals}
\label{sec:multicastinggoals}

In wireless networks, multicasting can have various performance goals. In particular, it is important to provide high quality of service (QoS), energy-efficient performance, and to ensure reliability and security. We discuss these performance goals next.

\vspace{1mm}
\emph{Quality of Service (QoS)}: A number of popular applications, such as teleconferencing, distance learning, and file caching and dissemination, exploit multicast transmissions while requiring a certain minimum QoS for satisfactory performance. To support such applications, providing network support for scalable group-based applications with strict QoS requirements becomes essential. Various metrics (such as end-to-end delay, jitter, loss, throughput, etc.) can be used to quantify the quality of service. The interested reader is referred to the survey \cite{striegel2002survey} for more details about the issues relating to QoS-based multicasting.

\vspace{1mm}
\emph{Power Consumption}: In wireless networks, the choice of the transmitting power level is linked with a crucial tradeoff between the reach and interference of a particular transmission. If the transmit power is increased, more of the neighboring receivers can be reached in a single-hop. This, however, comes at the cost of increased interference to other neighboring nodes that are not the intended receivers \cite{wieselthier2000construction}. In a CRN, the secondary nodes have to ensure that their transmissions do not interference with any PU. Another consideration impacting the choice of transmit power deals with node mobility as commonly mobile nodes have limited power resources thus motivating energy-efficient multicasting approaches. An efficient multicasting protocols must take into account all these issues. As an example work,
Thomas et al. presented a cognitive energy efficient multicasting approach that increases the lifetime of a multicast flow
by controlling the antenna directionality, transmission power, and routing tables of network nodes \cite{thomas2007cognitive}.
A comprehensive survey on energy efficient wireless multicasting protocols can be seen at \cite{gupta2005energy}.

\vspace{1mm}
\emph{Reliability and Security}: Ensuring reliability and security of multicast routing is complicated by the broadcast nature of wireless channels and the multicast group dynamics (through which nodes can join and leave groups at any time) \cite{akyildiz2006next}\cite{moyer1999survey}. It has been pointed out in literature that ARQ-based mechanisms, which require ACKs to be sent in response to a valid received packet, do not scale well for multicasting and broadcasting applications. Ensuring reliability and security is an important goal for multicasting protocols in CRNs and must be carefully considered by protocol designers.

\subsection{Classification of multicasting approaches}
\label{sec:classificationmulticasting}

In this subsection, we are going to classify multicasting approaches according to \emph{i)} the way multicast receivers are grouped and \emph{ii)} the way forwarding structure used for multicasting is structured. We discuss each of these in turn next.

\subsubsection{\emph{Grouping of the hosts}}

The conception of multicast group is an essential part of the overall multicasting framework. A multicast group can be \emph{static} or \emph{dynamic}: in the former category, the group composition cannot be changed once decided, while the latter category allows members to be added or removed at any time. The task of routing is more challenging for the case of dynamic groups since it is not known in advance which nodes will be added or removed \cite{oliveira2005survey}.

Multicast groups can be also be classified according to the relative number of users: in \emph{sparse} groups, the number of participants is small compared to the number of nodes in the network, while for \emph{pervasive} or \emph{dense} groups, most of the network nodes are engaged in multicast communication.

Alternatively, groups can be classified on the basis of its \emph{openness}: \emph{closed} group allows only group member senders to send data, while in an \emph{open} group the sender may or may not be a member.

Finally, groups can be classified on the basis of their permanency:
\emph{permanent} groups are everlasting and exist longer than transient groups \cite{deering1988host}.

\subsubsection{\emph{Forwarding structure based classification}}

Multicast protocols are broadly classified into the following two types.

\vspace{2mm}
\emph{1) Tree-based protocols}: In graph-theoretic terminology, a \emph{tree} on a graph is defined as a subgraph.  Tree-based multicasting protocols work by constructing a tree on the overall connectivity graph which connects together the multicast group in an acyclic subgraph. In a tree structure, each node can reach any other node through a single path. Trees have been widely deployed in wired networks for establishing shortest path routing: with such a tree being called a shortest-path-tree (SPT). Although they ensure efficient data forwarding for wireless networks, they are problematic for unreliable wireless networks due to lack of robustness associated with the fragile tree structure. In case of a link failure, a tree has to be recomputed because there is no alternative path present. Trees are well-suited for networks with relatively static topology (such as wired networks, or wireless mesh networks to some extent), but are not as suited for dynamic topologies (such as mobile ad-hoc networks or MANETs, and CRNs). Tree-based protocols can be further classified into source-based and shared-based tree protocols \cite{paul2002survey}.

There are fundamentally two distinct approaches to construction of multicasting trees: using \emph{i)} `shortest path trees' (SPTs), or \emph{ii)} minimum cost trees (MCTs). The former approach minimizes the distance of each receiver from the sender, while the latter approach aims to minimize the tree's overall edge cost \cite{nguyen2008multicast}.

The \emph{Source-based} approach works by constructing a \emph{SPT} rooted at the source node. Since the source-based trees are custom built for each source, the receivers in the multicast tree typically receive excellent QoS (as the receivers are connected to the source via shortest paths) at the cost of the usage of extra network resources. The problem with source based trees is mainly in its lack of scalability to large networks.

The \emph{Core-based} approach, on the other hand, typically uses \emph{MCT} algorithms such as the minimum Steiner trees (MSTs) to minimize the overall edge cost of the multicast tree. In the core-based tree (CBT) approach, a single tree is constructed for each group, regardless of the multicast source \cite{ballardie1993core}, to address the scalability problem faced by source-based trees. The core router receives all the messages as unicast transmission, and then forwards packets to all the ongoing interfaces of the tree except for the incoming one. CBT is able to conserve network bandwidth since it does not utilize flooding. The CBT is also able to scale to large networks since it utilizes a shared tree that has limited overhead. The drawback of a CBT approach is that since the tree is not customized to each source, the receivers will typically have non-optimal routes and QoS.

\vspace{2mm}
\emph{2) Mesh-based protocols} improve upon the fragility of tree-based structures by allowing multiple paths to exist between the source and the receivers. Mesh-based protocols avoid recomputation of trees on the failure of a link (which may be due to node mobility or PU arrival) and can instead rely on alternate paths that are part of the original forwarding structure. This introduces some robustness to link failures making mesh-based protocols more suited to wireless networks with dynamic topologies and network conditions (such as MANETs and CRNs). The drawback of mesh-based protocols is the higher computational and message overhead involved with managing the forwarding structure \cite{sahasrabuddhe2000multicast}.

\section{Challenges in Efficient Multicasting in CRNs}
\label{sec:challengesmulticasting}

To fully exploit the benefits of multicasting in CRNs, it is important to surmount the considerable challenges that accompany it. In this section, we divide the challenges into general challenges of multicasting, challenges due to wireless networks and finally challenges imposed by the cognitive environment.

In CRNs, whenever a PU starts communicating in the vicinity of a SU, it is required to vacate the channel and occupy another idle channel available, making the network highly dynamic. Therefore, deployment of multicasting becomes even more challenging with channel availability being dependent on both time and location \cite{almasaeid2010assisted}. This results in different sets of idle channels available at various SUs, making transmission coordination extremely exigent.

\subsection{Challenges unique to multicasting}

Optimal solution to the problem of multicasting is considered difficult even in centralized conditions. For example, the problem of optimally packing Steiner trees to find maximum multicast flow is NP-hard \cite{hodgskiss2010optimisation}.

In \emph{dynamic multicast groups}, nodes can dynamically subscribe or unsubscribe to a group. This leads to highly dynamic topology which creates problem due to the need of recomputation of the multicast forwarding structure. It is desirable that the group dynamics should not affect the way data is delivered currently to members that remain in the group \cite{diot1997multipoint}. Although, calculating an optimal multicast route is already a difficulty problem, maintaining route optimality after changes in the group and network complicates it further significantly \cite{diot1997multipoint}.

\subsection{Challenges due to the wireless medium}
\label{sec:wba}

Broadly speaking, there are three main challenges posed by wireless networks to the problem of efficient multicasting. We will discuss these next.

\subsubsection{\emph{Wireless broadcast advantage}}

Unlike in wireline networks, a transmission in wireless medium reaches all nodes in the transmission range \emph{simultaneously} assuming omnidirectional antennas. This is known as the \emph{`wireless broadcast advantage'} (WBA) or the \emph{`wireless multicast advantage'} (WMA) \cite{wieselthier2000construction}.

While exploiting WMA can result in lesser number of distinct transmissions than what would be required in an equivalent wired network, WMA is not really an advantage when it comes to devising polynomial-time optimal algorithms. In fact, the multicast `advantage' makes the multicasting problem much harder. As an example, the minimum energy broadcast problem, which is tractable and is solved easily for wireline networks through any of a number of algorithms devised for calculating the minimum spanning tree (MST), becomes NP-complete for wireless networks \cite{liang2002constructing} \cite{lun2005efficient}. We reiterate here that the broadcasting problem is a special case of the multicasting problem (where the multicast group contains all the network nodes). This implies that the multicast problem is at least as complex the broadcast problem, thereby motivating interest in development of heuristic multicasting algorithms \cite{wieselthier2000construction}.

\subsubsection{\emph{Interface diversity}}

Many wireless technologies can currently support provisioning of multiple radio interfaces on the same node. The use of multi-radio interfaces introduces `\emph{interface-diversity}' which allows multiple simultaneous transmissions on a node through these multiple interfaces tuned to orthogonal channels. It has been shown in previous work that exploiting the radio diversity can significantly improve routing performance both in the case of unicast \cite{draves2004routing} and broadcast \cite{qadir2006minimum}.

\subsubsection{\emph{Rate diversity}}

Another feature of modern wireless technology is the use of adaptive modulation to provide the ability to transmit at multiple link-layer transmission rates. As an example, IEEE 802.11b nodes support transmission at link-layer rates of 1, 2, 5.5 and 11 Mbps. This \emph{`rate-diversity'} provides an extra degree-of-freedom that should be incorporated into the design of multicasting framework. Although it is commonly assumed that multicasting and broadcasting will only utilize the lowest link-layer rate for transmission (with which maximum neighboring nodes can be reached), previous work has shown the benefit of using rate-diversity in the choice of link-layer transmission rate for both multicast \cite{qadir2006exploiting} and broadcast \cite{chou2006low}.

\subsection{Challenges introduced by CRNs}

Apart from the challenges posed by wireless networks in general, CRNs, in particular, pose the following three additional challenges:

\subsubsection{\emph{Channel diversity}}

In order to perform transmission coordination in CRNs, a common control channel (CCC) is commonly used for exchanging control information between SUs. The CCC design originates from multi-channel wireless networks, but when deployed in cognitive networks, it needs to address a number of additional challenges such as ensuring: robustness to PU activity, sufficient coverage of the CCC, and security of the CCC \cite{katabi2007trading}. Unless, there is a dedicated radio interface tuned to the CCC, the problem of `deafness' can arise and lead to coordination problems. A detailed discussion of issues relating to the use of CCC in CRNs is provided in \cite{lo2011survey}.

\subsubsection{\emph{Spectrum heterogeneity}}

The random and arbitrary nature of PU arrivals on various licensed channels can significantly complicate the problem of routing in CRNs. This implies that the spectrum available to a given SU is contingent on the activity of PUs at the given time in its range. The spectrum availability is likely to be dynamic (i.e., time-varying) and heterogeneous (i.e., not the same channels are available at all nodes) across the SUs in CRNs \cite{sengupta2013open}. Also, unlike traditional wireless networks the channels used are typically from the same spectrum band, it is possible that CRNs are operating on heterogeneous channels from diverse spectrum bands having distinctly different transmission properties. \cite{sengupta2013open}.

\subsubsection{\emph{Spectrum mobility}}

Arrival of a PU at a particular channel causes the SU using that channel to terminate its communication, and to find another path if available by switching to another channel. Frequent PU arrivals can lead to frequent temporal connection losses for SUs thereby seriously impacting their performance.

\section{Algorithms used for Wireless Multicasting}
\label{sec:algorithm}

Multicasting has been an active area of research for many decades and numerous techniques have been proposed for the construction of multicast routes. In a classic paper,
Diot et al. \cite{diot1997multipoint} described the basic multicasting algorithms in three categories: source-tree based, center-tree based, and Steiner-tree based algorithms \cite{oliveira2005survey}. These techniques are discussed next.

\subsection{Source Based Algorithms}

For \emph{source-based routing}, the routing tree created for each multicast group is rooted at the source node. Some implementations of this scheme utilize the reverse path forwarding (RPF) algorithm to ensure loop-free forwarding of multicast packets. The RPF algorithm has been shown to perform poorly for small multicast groups \cite{oliveira2005survey}. Source based routing techniques can also utilize Steiner tree-based methods to focus on minimization of tree cost \cite{oliveira2005survey}.

\subsection{Center Based Algorithms}

In contrast to source-based routing, \emph{center-based} tree, or the core-based tree, (CBT) algorithms utilize single unique shared tree for all group communication regardless of the source. The center-based algorithms constructs a tree rooted at a special root node known as the the \emph{center} node, or the \emph{core} node, through which all the group communication is managed. The center node is computed to have some special properties such as: e.g., being the closest to all other nodes \cite{oliveira2005survey}. Contrary to shortest-path-trees (SPT) used for source-based trees, the path between two group members in a CBT is not guaranteed to be the shortest. Furthermore, calculating the center node is a hard problem, although approximations have been proposed \cite{oliveira2005survey}.

\subsection{Steiner Tree Based Algorithms}

A \emph{Steiner tree} is a tree that interconnects a particular subset of the vertex set with minimum aggregate cost. While the Steiner problem for connecting a source with a multicast group apparently looks identical to the minimum-spanning-tree problem involving the same set of nodes, the difference is that the Steiner tree can incorporate other non-multicast-group vertices (the so-called Steiner points) and edges to produce the shortest interconnect between the source and the multicast group.

The Steiner tree is well-known to be a computationally complex problem and has been shown to be NP-complete \cite{winter1987steiner}. Also, since the edges are undirected, it is applicable only if we assume symmetric links. Another inefficiency is that it is a monolithic algorithm. Unlike incremental algorithms, monolithic algorithms have to be run every time there is a change in topology or costs. Although, it is aimed at a centralized computation, heuristic implementation can be distributed \cite{diot1997multipoint} \cite{bezenvsek2013survey}. Due to its complexity, a lot of attention has focused on deriving approximate solutions to this problem \cite{takahashi1980approximate} \cite{waxman1988routing} \cite{hauptmann2013compendium}.

The problem of routing with static multicast groups in \emph{wired} networks is often formulated as a Steiner tree problem \cite{oliveira2005survey}. In a wired network, multicast packets are transmitted over the tree edges independently (i.e., each outgoing link requires an independent transmission). Thus, the Steiner tree (which minimizes the overall aggregate cost of all the tree's independent links) is an adequate model for wired multicast.

\emph{Extension of Steiner trees to wireless networks:} Due to its characteristic WBA, the wireless multicasting problem is significantly different from the traditional multicasting problem in wired networks, and the wired multicasting approaches cannot be directly applied to wireless networks. For wireless multicasting, the \emph{Steiner connected dominated set} (SCDS) model has been proposed as a viable approach \cite{torkestani2011weighted}. The minimum SCDS problem was first proposed as a generalization of the well-known concept of a connected dominating set (CDS) by Guha and Khuller \cite{guha1998approximation}, and was applied to the problem of multicast in ad-hoc networks by Wu et al. \cite{ya2004construction}.


\section{Techniques used for Wireless Multicasting}
\label{sec:techniques}

Various techniques from diverse fields have been used in the design and analysis of wireless multicast. In this section, we survey the landscape of these techniques. In particular, we will discuss network-coding, optimization, machine-learning, and game-theoretic techniques in sections \ref{sec:netcoding}, \ref{sec:optimization}, \ref{sec:machinelearning}, and \ref{sec:gametheory}, respectively.

\subsection{Network Coding}
\label{sec:netcoding}

While routing was long viewed as the only primitive available for successfully operating multi-hop networks, it has been discovered that allowing intermediate nodes to diversify beyond the simple primitive of store-and-forward routing can actually improve performance. This discovery was first articulated in the seminar work of Ahlswede et al. \cite{ahlswede2000network}. This paradigm, called  network coding (NC), equips intermediate nodes with the ability of combining incoming packets and applying algebraic operations on received data before forwarding. The receiving node is then presented with a set of network coded packets as well as the information required for retrieving the original information. Network coding is considered a very attractive alternative to routing in settings like application-layer overlay networks and multi-hop wireless networks \cite{lun2006minimum}.

\begin{figure}[t]
\begin{center}
\includegraphics[width=.45\textwidth]{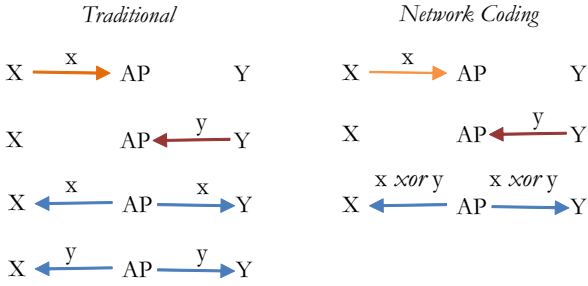}
\caption{Comparison between the traditional approach and the NC coding approach. It can be seen that even in this simple wireless networking topology of two nodes X and Y interconnected through an access point AP that the NC-based approach can reduce a transmission, and thereby avoid wastage of bandwidth, by exploiting WBA.}
\label{fig:nc}
\end{center}
\end{figure}

To develop an intuition for how network coding works, we refer the reader to fig. \ref{fig:nc}  representing a simple wireless network in which two nodes X and Y are connected through an access-point (AP)\footnote{While the advantage of NC over store-and-forward in wireless is often explained through the so-called ``butterfly'' example \cite{yeung2008information}, we use a simple example adapted from \cite{fragouli2006network}\cite{katti2006xors} to highlight both the NC advantage in avoiding redundant transmissions and also how it utilizes the WBA.}. Both the end nodes X and Y intend to send their packets, named x and y respectively, to the other node. In the traditional approach, this will require four transmissions as shown in fig. \ref{fig:nc}. NC-based approach, however, exploits the WBA to suppress an extra transmission by transmitting a single coded packet (determined through an XOR operation on X and Y). Due to the WBA, this packet is received by both the nodes X and Y, and since these nodes already known their own respective packet contents, they can easily recover the other packet. While we have used XOR operation as the coding choice at intermediate node for ease of exposition, in general, more elaborate algebraic coding operations (e.g., linear network coding) can be used to recombine several input packets into one or more output packets \cite{fragouli2006network}. Intuitively, NC improves network efficiency through extra computation at the end nodes. This tradeoff is appealing due to the increasing need of efficient network bandwidth and since processing continues to become cheaper and powerful riding the Moore's law.

While NC was originally proposed for multicasting in a wireline setting, it turns out that wireless applications are even more amenable to improvement with a NC scheme. This stems from the observation that NC naturally thrives when dealing with two characteristics of wireless links (its unreliability and its broadcast nature) that complicate the wireless routing problem \cite{deb2005network}\cite{lun2005efficient}. Interestingly, NC is useful for multicasting application both in lossless and lossy wireless settings \cite{lun2005efficient}. It was shown that for lossless networks (i.e., idealized networks with broadcast links that are effectively lossless), NC can achieve the maximum possible \emph{multicast} rate. Since NC allows joint use of network resources by multiple sink nodes, any rate possible for all sinks individually is simultaneously achievable for all sinks together \cite{ho2008network}. In lossless setting, the main utility of NC over routing is for multicast, while for single unicast sessions, NC provides no extra benefit. For lossy networks, NC, particular \emph{random linear NC}, provides better performance than routing for both single unicast and multicast sessions \cite{lun2005efficient}. It has been shown that for a given coding subgraph, random linear NC achieves the capacity of a single connection (unicast or multicast) \cite{ho2008network}.

NC can be categorized into the following two main types:

\vspace{1mm}
\emph{Intra-session NC} considers network coding for a \emph{single communication session} of either unicast communication to a single sink node, or multicast communication with multiple sink nodes. Our discussion up to now has focused on intra-session NC. Since in intra-session NC, the same set of sink nodes are decoding all coded packets, it suffices for each node to generate outputs through linear combinations of its inputs. The receiver can successfully decode the packets once it has received enough independent linear combinations of the source processes. Intra-session NC is well-understood with the maximum achievable multicast rate being characterized by the min-cut max-flow theorem \cite{ahlswede2000network}. The throughput benefits of intra-session NC are available only for multicast and vanish for the case of unicast. It has been shown that linear codes are sufficient for achieving maximum bounds for multicast traffic \cite{li2003linear} with polynomial-time linear encoding and decoding algorithms being available \cite{koetter2003algebraic}.  Also, randomized NC algorithms focusing on multicast have been proposed \cite{ho2003benefits}.

\vspace{1mm}
\emph{Inter-session NC} (i.e., coding amongst information symbols of different sessions) is needed in general for achieving optimal rates when multiple sessions share the network.  Inter-session admits throughput benefits for the cases of both unicast and multicast, with these benefits readily visible in two special kinds of network topologies: those having ``butterfly'' substructure or a ``wireless cross'' topology \cite{khreishah2009cross} (similar to the topology used in figure \ref{fig:nc}).
Inter-session NC is typically more complicated than intra-session NV requiring strategic coding so that each sink may be able to decode its desired source processes. Another complication is that in inter-session NC, decoding may be required at non-sink nodes unlike the case of intra-session NC. Presently, it is not yet completely known how to construct optimal network codes for multi-session network problems, although some work has been done in this direction \cite{ho2006random}.

It was discussed earlier (section \ref{sec:wba}) than minimum-energy multicasting is an intractable problem if we consider a routing based solution. This is because constructing the minimum-cost multicast tree, in traditional routed networks, requires solving the directed Steiner tree problem which is known to be NP-complete. However, the situation is not as bleak for NC based multicast solution as the problem of finding the minimum-cost multicast tree, when network coding is used, can be solved in polynomial time with a \emph{linear program}. The results of this optimal NC-based solution greatly improves the results of non-NC-based routing heuristics \cite{lun2005efficient}. Furthermore, the practical implementation of NC is eased as decentralized NC solutions exist crucially in wireless networks. The amenability of NC for decentralized implementation, more than the energy savings, arguably holds the most promise for improving the performance of wireless networks \cite{lun2005efficient}.

Salient benefits of NC can be summarized to include \emph{i)} increase in throughput, \emph{ii)} improved robustness, and \emph{iii)} lesser complexity \cite{ho2003benefits}. The most well-known utility of NC is \emph{increase in throughput}. Ahlswede et al. \cite{ahlswede2000network} demonstrated an increase in throughput for the problem of multicast in a wireline network. This observation holds even more true for the same problem in wireless settings, where NC provides a pronounced advantage and yields a throughput advantage over routing \cite{ho2008network}. Another important utility of NC is that it provides \emph{improved robustness}: implying loss resilience and facilitation in the design of localized distributed algorithms that can perform well even with partial information \cite{fragouli2006network}. While, erasure channel coding is traditionally performed at the source node, it has been shown that NC is useful in combating against packet losses leading to improved error resilience \cite{ho2008network}. Finally, NC often leads to \emph{less complex solutions} as compared to the routing approach.
As an example, the routing problem of constructing a Steiner multicast tree which selects a minimum-cost subgraph is computationally complex (i.e., it is in the NP-complete class) even in a centralized setting. The corresponding NC problem, however, is considerably less complex and can be solved through linear optimization with low-complexity distributed solutions \cite{ho2008network}.

In an empirical study, it was shown on a 20-node wireless network that a simple network coding based approach (that uses the XOR operation to combine packets) significantly improves performance \cite{katti2006xors}. In another work, the interaction between \emph{network coding} and \emph{link-layer transmission rate diversity}\footnote{The ability of a modern wireless node to support multiple link-layer transmission rate: e.g., IEEE 802.11b supports rates of 1, 2, 5.5 and 11 Mbps.} in \emph{multi-rate} multi-hop wireless networks has been investigated by Vieira et al. \cite{vieira2013fundamental} where it was shown
that network coding can be combined with multi-rate link layer broadcast to increase network throughput for multicast applications.

\subsection{Optimization}
\label{sec:optimization}

Optimization theory is a richly developed theory comprising tools and techniques for determining ``\emph{optimal}'' decisions in scenarios which may also incorporate certain constraints \cite{keshav2012mathematical} \cite{hillierintroduction}.

Formally, a mathematical optimization problem has the following form:

\begin{equation}
\mbox{ minimize } f_0(x)\\
\mbox{ subject to } f_i(x) \le b_i
\end{equation}

Here the vector $\mathbf{x}$ = $(x_1,  \dots, x_n)$ is called the \emph{optimization variable}, and the function ($f_0 : \mathbf{{R_n}} \rightarrow \mathbf{R}$) of the optimization variable, that we have the objective of minimizing, is known as the \emph{objective function}. The functions $f_i: \mathbf{R_n} \rightarrow \mathbf{R}$, $i = 1, \dots,m$ are the (inequality) \emph{constraint functions}, and the constants $b_1, \dots, b_m$ are the limits, or bounds, for the \emph{constraints}. A vector $\mathbf{x}^*$ ``\emph{solves}'' the optimization problem, or is deemed \emph{optimal}, if it has the smallest objective function value among all vectors that satisfy the constraints defined.

There are many classes of optimization problems generally characterized on the basis of the form of the objective and the constraint functions. In particular, for \emph{linear program} (which we will study in section \ref{sec:linearprogramming}), the objective function $f_0$ and the $m$ constraint functions $f_1, ...., f_m$ are all linear: i.e., $f_i(\alpha x +\beta y) = \alpha f_i(x) + \beta f_i(y))$. If the optimization problem is not linear, it is called a \emph{nonlinear optimization} problem. The class of \emph{convex optimization} (which we shall study in section \ref{sec:convexopt}), which includes linear optimization as a special case, the objective function $f_0$ and the $m$ constraint functions $f_1, \dots, f_m$ are all convex: i.e., $f_i(\alpha x +\beta y) \le \alpha f_i(x) + \beta f_i(y))$. Due to the inequality in the preceding constraint function, convex programming can be of both linear and nonlinear types.

An important characteristic of an optimization problem is its \emph{discrete} or \emph{continuous} nature. Typically, continuous optimization problems either have no constraints or have constraints of a continuous character comprising equations and inequalities.  In problems of discrete optimization, also called \emph{combinatorial optimization}, either the constraint set is finite or it has a discrete nature. Informally speaking, combinatorial algorithms are techniques for high speed manipulation of combinatorial objects such as permutations, graphs, and networks \cite{knuth2006art} \cite{papadimitriou1998combinatorial}. Common applications of combinatorial optimization include scheduling, assignment, route planning, set covering, etc. We will find extensive use of combinatorial optimization in multicasting \cite{leggieri2009multicast}. Two important combinatorial optimization techniques, which are extensively used in multicasting literature, are \emph{linear programming} (covered in section \ref{sec:linearprogramming}) and \emph{integer programming} (covered in section \ref{sec:IP}). Linear programming is a problem of combinatorial optimization according to its fundamental theorem, while integer programming is a combinatorial technique in which optimization variables can only adopt discrete integer values \cite{papadimitriou1998combinatorial}.

Optimization problems come in two main varieties if we consider the \emph{complexity of the optimization problem}. Firstly, there are some ``easy'' problems that can be solved in time bounded by a polynomial in the input length $n$, and secondly, a larger class of ``hard'' problems for which no polynomial time algorithm exists and all known algorithms require time exponential in $n$ in the worst-case \cite{fisher2004lagrangian}. If a problem has a linear objective function and linear constraints, the problem is considered `easy' to solve and linear programming (discussed in section \ref{sec:linearprogramming}) can be used to solve it (even for very large sized problems). For many practical problems, the objective function or the constraints are not linear, and the resulting problem is `hard' to solve in a time-efficient manner. A linear programming model is an inadequate match for such systems, and other approaches are necessary. We will discuss various approaches that can be used to tackle such `hard' problems later on.

Deriving a \emph{solution to an optimization problem} typically entails seeking an optimal solution (if the problem is tractable or is a small-scale NP problem), or an approximation solution with some qualification of its quality in comparison with the optimal, or aiming for heuristic if even an approximate solution is difficult to obtain. In case of multiple objective functions, one way of defining efficiency of solution is through \emph{Pareto optimality} which captures the tradeoff between the multiple objective functions by choosing an operating point where further improvement in any objective function cannot be made except through deterioration in some other objective function's value.

\begin{figure}[t]
\begin{center}
\includegraphics[width=.4\textwidth]{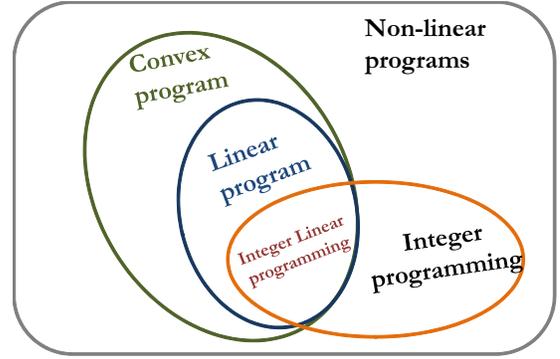}
\caption{Relationship between various optimization approaches. Convexity is the ``watershed'' between tractable and intractable problems. Linear programs generally have polynomial-time solutions while Integer programs have no known polynomial-time solutions as they belong to the class of NP-complete problems.}
\label{fig:opt}
\end{center}
\end{figure}

In \emph{optimization-based works specific to multicasting}, a cross-layer optimization framework for multi-hop multicast in wireless mesh networks was proposed in \cite{yuan2006cross}. A survey of optimization problems that are relevant to multicast tree construction are presented in \cite{oliveira2006optimization} \cite{oliveira2005survey}.

\emph{Optimization goals in multicast: }Different objectives can be considered when optimizing a multicast routing problem, such as, for example, path delay, total cost of the tree, and maximum congestion \cite{oliveira2005survey}. Some important optimization problems that relate to multicast routing are the ``the multicast network dimensioning problem'', and the ``multicast packing problem'', \cite{oliveira2005survey}. The problem of optimally packing Steiner trees to find maximum multicast flow is NP-hard \cite{hodgskiss2010optimisation}. This was shown in the context of single-radio single-channel multi-hop wireless network in  \cite{wan2009multiflows} where it was shown that the `maximum multiflow' and the `maximum concurrent multiflow' problems subject to bandwidth and interference constraints are NP-hard.

The various \emph{applications of optimization techniques to telecommunications} are surveyed in detail in the handbook \cite{resende2006handbook}. In recent times, there have been efforts in designing optimizable networks \cite{chiang2007layering} \cite{palomar2006tutorial} and protocol design for communication networks as a distributed resource allocation problem \cite{shakkottai2008network}.

\subsubsection{\emph{Linear Programming}}
\label{sec:linearprogramming}

Linear programming (LP)\footnote{The use of the `programming' does not refer to computer programming, but is used in the sense of planning; this word is used in the same sense in mathematical programming, linear programming, and dynamic programming.} is one of the biggest success stories of `optimization theory'. LP techniques have been exploited by generations of practitioners to solve practical problems in a diverse set of fields such as engineering, economics and finance \cite{hillierintroduction} \cite{bertsimas1997introduction}. LP is commonly applied in these fields to realize ``optimal'' logistical planning and scheduling. An application area of LP, much closer to our subject, is in \emph{network optimization}.

Typical network optimization problems, that may be formulated as linear programming problems, are the shortest path problem, the min-cut max-flow problem, and the minimum cost-flow problem  \cite{ahuja1993network}. Various highly efficient solutions have been devised that exploit the special structure found in minimum cost network flow problems. This structure allows radical simplification to the standard Simplex solution method of LP and allows even large-scale optimization problems to be solved very efficiently. The  foundation of these procedures can be seen in a specialized field of optimization that caters to network flow problems \cite{ford2010flows} \cite{ahuja1993network}.

One of the most important discoveries in the early development of linear programming was the concept of \emph{duality} and its many important ramifications. This discovery revealed that every linear programming problem has associated with it another linear programming problem called the \emph{dual} \cite{hillierintroduction}. The relationships between the dual problem and the original problem (called the primal) prove to be extremely useful in a variety of ways \cite{hillierintroduction}.

\vspace{2mm}\emph{LP Solutions}: Historically, the first solution proposed for linear programming was the \emph{Simplex algorithm} proposed by Dantzig in 1948. In mere terms of its widespread application, the \emph{``simplex algorithm''}, proposed by George Dantzig, is arguably the most successful algorithm of all time.\footnote{The simplex algorithm was named as the one of the top ten algorithms of the 20th century \cite{cipra2000best}.} Decades of refinements have led to very efficient forms of the simplex algorithm that can routinely solve large-scale linear programs with hundreds of variables and thousands of constraints. While its performance for typical programs is quite efficient,  the simplex algorithm performance in the worst-case (for specially crafted pathological examples) is poor requiring exponential number of steps \cite{papadimitriou1998combinatorial}.

The state-of-the-art in LP solutions was further extended in 1984 when Karmarkar published a classic paper \cite{karmarkar1984new} on a technique which has become known as the \emph{interior-point method}. The \emph{primal-dual} algorithms are the most important and useful algorithms from the class of interior-point methods and possess excellent theoretical and practical properties \cite{wright1997primal} that apply widely beyond the class of linear programming (e.g., it is applicable to convex programming). It was proposed by Dantzig, Ford, and Fulkerson as an alternative method of solving linear programs \cite{goemans1997primal}. In the primal-dual method for approximation algorithms, an approximate solution to the problem and a feasible solution to the dual of an LP relaxation are constructed simultaneously; the performance guarantee is proved by comparing the values of both solutions \cite{goemans1997primal}. It can be used to provide good approximation algorithms for a wide variety of NP-hard problems \cite{goemans1997primal}\cite{karimi2010multicast}.

\subsubsection{\emph{Integer Programming}}
\label{sec:IP}

In many optimization problem, it only makes sense for certain optimization variables to take on integer values (e.g., number of packets, flows, etc. generally make sense for integral values only). When all the optimization variables must take only integer values, the optimization model is known as integer programming (IP). In general, IP problems can belong to either of the linear or the nonlinear class. If an IP belongs to the linear class, i.e., its objective and constraint functions are both linear, the class of model is referred to as integer linear programming or ILP. If in an optimization model, certain optimization variables can only take on integer values while other can take real values, the class of optimization model is known as mixed integer programming (MIP) model. The relationship of IP with other classes of optimization is depicted graphically in figure \ref{fig:opt}.

While LP problems generally entertain polynomial-time solutions, IP programs are more complex to solve and typically are in the class of non-deterministic polynomial-time (NP). Since IP problems are computationally intractable (i.e., they are NP-complete or NP-hard), various \emph{relaxation techniques} have been used for producing approximate solutions. IP models can be solved more efficiently if the problem has network substructure in which case \emph{Lagrangian relaxation} can be used to decompose the IP. Other solution concepts for IP include \emph{branch-and-bound} and \emph{branch-and-cut}. The most common relaxation method used for solving IP, though, is the linear programming relaxation through the restriction of optimization variables taking integer values is relaxed.

IP techniques are useful in communication networks for \emph{synthesis}, \emph{assignment} and \emph{scheduling} problems \cite{resende2006handbook}.  The objective in a synthesis problem is to find an optimal allocation of resources to edges of a complete graph such that all demands, and other requirements of the problem, are met \cite{resende2006handbook}.  An example of a synthesis problem is ensuring that the transmission demands are met even in the event of a node or link failure. IP techniques are also commonly used for solving a wide variety of assignment and scheduling problems \cite{resende2006handbook}. Since allocation often does not entertain a fractional solution, IP formulation arises as a natural candidate optimization model. Example applications include: \emph{i)} channel assignment problem which assigns channels to the various nodes and their interfaces, and \emph{ii)} allocating discrete time slots to job requests.

\emph{Applications to routing and CRNs:} Many practical problems in wireless networks can be formulated as IP problems, and IP has been popularly used for modeling the broadcast and multicast problems in wireless networks. As an example, Das et al.  presented integer programming models for the minimum power broadcast/ multicast problem in wireless networks \cite{das2003minimum}.

\subsubsection{\emph{Convex Optimization}}
\label{sec:convexopt}

Although, earlier it was thought that all nonlinear problems are intractable, it has been shown now that convexity defines the demarcation, or the ``watershed'', between tractable and intractable problems \cite{chiang2007layering}. This insight, along with the invention of efficient methods for solving convex optimization problems, marks the biggest progress in the field of optimization theory since the formulation of linear programming and invention of the simplex algorithm. So much so that it can be said that successfully formulating a practical problem as a convex optimization problem is as good as solving the problem \cite{boyd2004convex}.

Interest in convex optimization has been reinvigorated by a few important recent discoveries. It has been shown that interior-points methods developed for solving linear programming problems are useful for solving a broader much wider class of convex optimization problems as well. Secondly, it is now realized that convex optimization problems (beyond least-square and linear optimization problems) are much more prevalent than previously thought \cite{boyd2004convex}. Convex optimization subsumes least-square, linear-optimization, conic programming \cite{nemirovski2006advances}, and geometric programming \cite{chiang2005geometric} classes of optimization \cite{boyd2004convex}.

Formulating a problem as a convex optimization problem can be rewarding for two particular reasons. Firstly, a convex optimization problem can be reliably and efficiently solved through various special methods (e.g., the interior-point method proposed for linear optimization). Secondly, \emph{duality theory} often reveals an interesting interpretation of the primal problem and often leads to an efficient distributed method for solving it.

Convex programming is useful even for non-convex programming problems. In such a case, a non-convex problem can be approximated by an approximate convex model, solving which can give a starting point for a local optimization search applied to the original non-convex problem \cite{boyd2004convex}. Most practical methods of global optimization of non-convex problem rely on convex optimization in some way. In \emph{relaxation}, non-convex constraints are replaced with a looser convex constraint, while for the method of \emph{Lagrangian relaxation}, the Lagrangian dual problem, which is a convex problem, is solved to provide the lower bound on the optimal value of the non-convex problem \cite{boyd2004convex}.

\emph{Applications to routing and CRNs:} Two major applications of convex optimization has been network utility maximization and robust transceiver design. A survey of application of convex programming techniques to communications and signal processing is provided in \cite{luo2006introduction}.

\subsubsection{\emph{Dynamic Programming}}
\label{sec:DP}

Dynamic programming (DP) is a powerful optimization technique applicable when a problem can be decomposed into subproblems such that the optimal solution to the original problem is a composition of the optimal solutions to each subproblem. (This is also called the \emph{optimal substructure} of the problem.) The technique proceeds by decomposing the problem into smaller subproblems, solve each subproblem recursively, and then aggregating the solutions to obtain the final answer. DP is applicable where the problem can be divide into smaller problems that are easily solved and when the program has the optimal substructure which allows coalescing the smaller solutions into a solution of the overall problem. It is critical for DP that the recursion does not result in a proliferation of subproblems \cite{keshav2012mathematical}.

The term ``dynamic programming'' was used by Richard Bellman in the 1940s in his mathematical study of sequential multi-stage decision processes in which it is desired to make an ``optimal'' decision one stage after another. The term `dynamic' in `dynamic programming' refers to the \emph{temporal} aspect of multi-stage decision making while `programming' refers to optimization. Naively, one would assume that an optimal decision should require considering the set of all feasible policies, and then choosing the policy which provides the maximum return. This brute-force approach does not work except for trivial toy problems and is grossly inadequate for processes involving even a moderate number of stages and actions. The excessive dimensionality of such brute-force enumeration based approach results from trying to incorporate too much information into the framework.  The basic idea of the theory of DP is refreshingly simple. It proposes that `optimal policy' should be viewed as determining the decision required at each time in terms of the \emph{current state} of the system. This is known as the \emph{principle of optimality} which is based on the insight that the remaining decisions in DP must constitute an optimal policy $\pi$ for the continuation process treating the current state as starting input regardless of the initial states and decisions. Using this principle, the optimal policy can be computed through backward induction starting at the terminal point.

Since DP proceeds by breaking down a complex problem into smaller subproblems, it is important to for time-efficiency that each subproblem is solved only once. In many DP problems, it turns out that the repeating subproblems grow exponentially as a function of the size of the input. DP solves this problem by trading off space for time. In DP,  the solution to a subproblem is stored, or ``memoized'', and looked up when needed. DP combines the best of the methods of `greedy algorithms' and `exhaustive search' into an intelligent brute-force method that allows us to go through all the possible solutions to pick up the optimal solution in reasonable time.

Various famous algorithms are based on DP principles: e.g., the famous Viterbi algorithm, the Dijkstra algorithm, and the Floyd-Warshall algorithm all utilize DP ideas \cite{leiserson2001introduction} \cite{keshav2012mathematical}. In general, DP-based solutions are highly suited to problems that involve graph-theoretic problems involving rooted trees \cite{skiena1998algorithm}.

\vspace{2mm}
\emph{Application of DP to routing problems:}  Lun et al. formulated the problem of \emph{dynamic multicasting}, which aimed at finding minimum-cost \emph{time-varying} multicast subgraphs that provide continuous service to dynamic groups in wireless networks using \emph{network coding}, within the framework of DP \cite{lun2005dynamic}. We have seen earlier in section \ref{sec:netcoding} that minimum-cost multicast tree can be computed when network coding is used with a linear program---however, such works have assumed a static multicast group setup that does not change with time. The dynamic multicasting problem is more challenging and DP techniques have been applied to this problem with some success in \cite{lun2005dynamic}.

\vspace{2mm}
\emph{Application of DP in CRNs:} DP, like most other optimization techniques, can only be used in CRNs when perfect knowledge about the system is available.\footnote{For the case where perfect knowledge is not known, machine-learning techniques like reinforcement learning are more appropriate. Such machine-learning techniques are covered later in section \ref{sec:machinelearning}.} In such a scenario, i.e., with a model of the environment available, the wireless environment is typically modeled as a Markov decision process (MDP) which provides a mathematical framework for modeling sequential planning  by an agent in stochastic situations, where the outcome does not follow deterministically from actions, as an `optimal control' problem in which the aim is to select actions that maximize some measure of long-term reward. DP techniques are widely used for solution of MDPs.

\subsubsection{\emph{Decomposition methods}}

Decomposability techniques have been extensively used in optimization to lead to distributed (and often iterative) algorithms that converge to the global optimum. In wireless networks, distributed solutions are particularly attractive as a centralized solution may be non-scalable, too costly or fragile \cite{chiang2007layering}. Decomposition theory naturally provides the mathematical language to build an analytic foundation for the design of modularized and distributed control of networks \cite{chiang2007layering}. The method of decomposition is considered an extremely important versatile tool vital for practical distributed solutions of optimization problems. It has been stressed in \cite{chiang2007layering} that the importance of ``decomposability'' to distributed solutions is akin to the importance of ``convexity'' in efficient computation of global optimum.

The work in \cite{chiang2007layering} attempts to coalesce the mechanisms at various protocol layers into a unifying holistic theory which can be used to study the architectural and performance issues in protocol layering. It is assumed that the layered protocols are performing asynchronous distributed computation over the network to solve a global network utility maximization (NUM) problem implicitly. In this conception, different layers locally iterate on different subsets of the decision variables asynchronously to achieve individual optimality, with the localized algorithms then collectively achieving a global objective.

\subsubsection{\emph{Nonlinear optimization}}
\label{sec:nonlinear}

We have earlier seen in the beginning of section \ref{sec:optimization} that if the objective function ($f_0$) or the $m$ constraint functions ($f_1, ...., f_m$), then the optimization problem is known as nonlinear optimization.

For constrained non-integer linear optimization, or simply linear programming, it is well-known that the objective function has its maximum and minimum values at one of the vertices of the polytope defined by the intersection of the constraint planes. Since the extremal value of the objective function is guaranteed to be at a vertex of the polytope, linear programming can thus be easily solved through methods like the simplex algorithm. In nonlinear optimization, on the other hand, there is no such guarantee that the extremal value of the objective function will lie at a vertex of the polytope. In particular, the objective function may both increase and decrease as we walk along the contour line \cite{boyd2004convex}, thereby making the optimization problem much more challenging. Since the technique of checking the vertices of the polytope of constraint planes is no longer sufficient, we need to resort to any of the large number of nonlinear optimization techniques proposed in literature \cite{keshav2012mathematical}.

Nonlinear optimization techniques fall into roughly into two categories \cite{keshav2012mathematical}:

\begin{enumerate}

\item \emph{Constrained nonlinear optimization:} When the objective function and the constraints are continuous and at least twice differentiable, there are two well-known techniques: Lagrangian optimization and Lagrangian optimization with the Karush-Kuhn-Tucker (KKT) conditions.

\item \emph{Heuristic nonlinear optimization:} When the objective functions are not continuous or differentiable, we are forced to use heuristic techniques, such as hill climbing, simulated annealing, and ant algorithms.

\end{enumerate}

\vspace{1mm}
We discuss these two techniques in the next two subsections.

\subsubsection{\emph{Constrained nonlinear optimization}}
\label{sec:constrainednonlinear}

\vspace{2mm}
\emph{Lagrangian Optimization:}
\vspace{2mm}

Lagrangian optimization is a method of constrained nonlinear optimization that computes the maximum or minimum of a function $f$ of several variables subject to one or more constraint functions denoted $g_i$. The method of Lagrange multipliers gives the set of necessary conditions to identify optimal points of equality constrained optimization problems. This is done by converting the original constrained problem to an equivalent unconstrained problem with the help of certain unspecified parameters known as Lagrange multipliers.


As an example, consider that we wish to maximize $f(x, y)$ subject to the constraint that $g(x, y) = c$. It is assumed that both these functions have continuous first partial derivatives. The Lagrange multiplier $\lambda$ is used in the \emph{Lagrangian dual function} $L$, or simply \emph{Lagrangian}, in the following way: $L(x,y,\lambda) = f(x,y) + \lambda \cdot \Big(g(x,y)-c\Big)$. The Lagrangian can be solved to recover the solution to the original constrained optimization problem \cite{bertsekas1982constrained}.

Lagrange optimization is based on the two fundamental principles of \emph{duality} and \emph{relaxation}.

\begin{enumerate}

\item \emph{The duality principle} implies that an optimization problem can be viewed from either from the perspective of the \emph{primal problem} or the alternative perspective of the \emph{dual problem}.  The primal and dual problems are intimately linked with the solution of the dual problem informing about the solution of the primal problem. The optimal values of the primal and dual problems are not necessarily the same though, with the solution of the dual providing a lower bound \cite{boyd2004convex} to the solution of the primal problem. The difference between the solutions for the dual and the primal is known as the \emph{``duality gap''}. This duality gap is zero under a constraint qualification condition for convex optimization problems which ensures \emph{``strong duality''}. Thus, when the necessary condition is fulfilled, the value of an optimal solution to the primal problem establishing the immense utility of the duality principle. In summary, the main insight behind using duality is to bound or solve an optimization problem via a different optimization problem.

\vspace{2mm}
\item \emph{The Lagrangian relaxation} idea is based on the intuition of conceptualize a ``hard'' optimization problem (e.g., MILP) as an ``easy'' problem that is complicated by a relatively small number of complicating side constraints \cite{fisher2004lagrangian}. This idea proposes linking the original minimization problem, termed as the \emph{primal problem}, with a complementary maximization problem, termed as the \emph{Lagrangian dual problem}, which is often easier to solve as it readily presents decomposition possibilities. The basic intuition of Lagrange duality is to develop a relaxed version of the primal problem by transferring the complicating constraints to a modified objective function (known as the Lagrangian function or simply the Lagrangian) augmented with a weighted sum of the constraint functions in which the weighting term, representing the \emph{Lagrangian multipliers}), is used to penalize in proportion to the amount of violation of the dualized constraints. The Lagrange dual is obtained by minimizing the Lagrangian function with respect to the primal variable. Interestingly, solving the Lagrange dual is always a convex optimization problem, even when the primal problem is not. The solution of the Lagrange dual is either the lower bound for the primal solution (for weak duality) or the exact primal solution (if conditions of strong duality are fulfilled).
\end{enumerate}

\vspace{2mm}
\emph{Karush-Kuhn-Tucker Conditions:}
\vspace{1mm}

The Lagrangian method is applicable when the constraint function are limited to being equality constraints. Allowing inequality constraints, the Karush-Kuhn-Tucker (KKT) approach generalizes the Lagrange multiplier method to allow solutions to a broader class of nonlinear programming problems. Using the KKT conditions, it can be determined whether the stationary point of the auxiliary Lagrangian function is also a global minimum \cite{keshav2012mathematical}.

\emph{Applications to routing and CRNs:} The methods of Lagrangian relaxation and dual decomposition are popular techniques that are applied to a wide variety of optimization problems in wireless routing.  In \cite{yuan2006cross}, the throughput maximization problem for multi-hop multicast is decomposed into two subproblems: firstly, a data routing subproblem at the network layer, and secondly, a power control subproblem at the PHY layer. The coordination between these subproblems is managed through a set of Lagrangian dual variables. In \cite{chen2006optimal}, it is proposed that cross-layered design be performed systematically through the framework of ``layering as optimization decomposition'' \cite{chiang2007layering} for time-varying channels. The resource allocation problem is broken down, through dual composition, into three subproblems of congestion control, routing and schedule which interact through the congestion price. A recent Lagrangian relaxation based wireless routing work can also been seen at \cite{wen2011minimum}.

\subsubsection{\emph{Heuristic nonlinear optimization}}
\label{sec:heuristicnonlinear}

For the typical class of intractable problems, sometimes we are satisfied with heuristic or ``good enough'' solutions that provide reasonably good results most of the time. Amongst various heuristic solutions, we will discuss greedy algorithms, genetic algorithms, and ant colony optimization algorithms.

\emph{Greedy algorithms:} In a \emph{greedy algorithm}, the basic building block of a complete feasible solution is a partial solution developed by ``greedy'' that are based on whatever partial information is available at the time. These partial solutions are progressively developed to build a more complete solution until the iterations develop a complete feasible solution. The famous Dijkstra algorithm used for solving the shortest path tree (SPT) problem is an example greedy algorithm.

Randomization is an important tool that can be exploited to avoid local minima's in search-based optimization problems. In particular, randomization is utilized by tools like \emph{simulated annealing}. Another core idea adopted in some metaheuristic techniques like \emph{tabu-search} is to use \emph{adaptive memory} contrary to the approach adopted in memoryless approaches like simulated annealing.  Metaheuristics also employ concepts of \emph{intensification} (which encourages intensifying previous solutions found to perform well) and \emph{diversification} (which encourages search to examine unvisited solutions).  In some other fields (e.g., in genetic algorithms and reinforcement learning), the concepts of intensification and diversification are known by the terms \emph{exploitation} and \emph{exploration}, respectively.  It is to be noted that exploitation and exploration, or alternatively, intensification and diversification, represent conflicting goals and therefore the dilemma of choosing one or the other needs to be resolved in a balanced fashion.

There are various other metaheuristic techniques proposed in literature and the interested reader is referred to a book on this topic \cite{talbi2009metaheuristics} or the book chapter on this topic in \cite{resende2006handbook}. We discuss next two particular evolutionary algorithm metaheuristic algorithms: genetic algorithms and ant-colony optimization.

\vspace{2mm}\emph{Genetic algorithms:} A genetic algorithm (GA) is a particular class of evolutionary algorithm which uses techniques inspired from evolutionary biology---such as inheritance, mutation, crossover, and natural selection---to improve the performance of a computational process. In every generation, multiple individuals are stochastically selected from the current population with fitter individuals more likely selections and are genetically modified (mutated or recombined) to form the next generation of the population. The usage of genetic operators and stochastic selection allow a gradual improvement in the `fitness' of the solution and allow GAs to keep away from local optima. Various genetic algorithms based solutions have been proposed for wireless multicasting and these include \cite{roy2004qm2rp} \cite{cheng2010genetic}.

\vspace{2mm} \emph{Ant Colony Optimization:} While typical `shortest path' routing protocols may have significant computational and message complexity, the humble biological ants, in a marvel of nature, are able to shortest routes to food sources in the dynamics of ant colony with extremely modest resources. A lot of research effort has been focused on imitating the performance of biological ants to produce optimized and efficient distributed routing behavior in wireless networks \cite{zhao2012cognitive} \cite{di2005anthocnet} \cite{shen2008ant}. In particular, a few ant-colony-optimization based multicasting protocols have been proposed for wireless networks \cite{roy2004qm2rp} \cite{yi2011improved}.

\subsection{Machine Learning}
\label{sec:machinelearning}

\emph{Machine learning} is a interdisciplinary field that deals with learning systems and algorithms. It draws upon techniques and methods from a wide variety of fields such as statistics, information theory, artificial intelligence, optimization theory, control theory, operations research, etc.  \cite{russell1995artificial}. Russell and Norvig \cite{russell1995artificial} describe machine learning to be the ability to ``adapt to new circumstances and to detect and extrapolate patterns''. Machine learning techniques have proven themselves to be of great practical utility in diverse domains such as pattern recognition, robotics, natural language processing, autonomous control systems. They are particularly useful in domains, like CRNs, where the agents must dynamically adapt to changing conditions.

\subsubsection{\emph{Reinforcement Learning}}
\label{sec:reinforcementlearning}

The aim of an intelligent agent in reinforcement learning (RL) is to determine a \emph{policy}, or a sequence of actions, that maps the state of an \emph{unknown stochastic} environment to an optimal action plan. RL thus addresses the \emph{planning} problem for unknown stochastic environments. For the case where the environment is stochastic but with a known model (i.e., we have an \emph{unknown stochastic environment}), the framework of Markov decision processes (MDPs) is used instead. RL techniques are widely used in CRNs, with their importance emanating from the specific relevance of RL techniques to CRN environment (which is highly dynamic, unpredictable, and generally unknown a-priori).

Since the environment RL agents work in are stochastic, the payoffs of actions are also not deterministic. The agent, therefore, has to balance two potentially conflicting considerations as it performs sequential decision making. On the one hand, it needs to \emph{explore} all feasible actions and their consequences to determine which action returns the most value. Simultaneously, it is also desired to \emph{exploit} the existing knowledge, attained through past experience, of favorable actions which received the most positive reinforcement.

An interesting way to conceptualize the difference between RL and MDPs is think of RL as a simulation-based technique for solving large-scale and complex MDPs. Crucially, RL can solve MDPs without explicit specification of the transition probabilities. These values are needed by classical dynamic programming solutions of value and policy iteration. In addition, RL can work with very large number of states when used along with function approximation.

Most RL algorithms can be classified into being either \emph{model-free} or \emph{model-based} \cite{sutton1998reinforcement}. A model intuitively is an abstraction that an agent can use to predict how the environment will respond to its actions: i.e., given a state and the action performed therein by the agent, a model can predict the (expected) resultant next state and the accompanying reward. In the \emph{model-free approach}, which are most applicable to CRNs with dynamic unknown conditions, the agent aims to \emph{directly} determine the optimal policy by mapping environmental states to actions without constructing a MDP model of the environment. An example of a popular model-free RL technique is the Q-learning technique whose application to multicast routing in CRNs we will study next \cite{sutton1998reinforcement}.

\vspace{2mm}
\emph{Q-learning}: Q-learning is a popular model-free RL technique with limited computational requirements that enables agents to learn how to act optimally in controlled Markovian domains. The implication of being model-free is that Q-learning does not explicitly model the reward transition probabilities of the underlying process. Q-learning proceeds instead by estimating the value of an action by compiled over experienced outcomes using an idea known as \emph{temporal-difference (TD) learning}. The TD learning idea has been referred to as the central key idea in the theory of RL. TD learning combines ideas from \emph{Monte Carlo (MC) methods} and dynamic programming (DP). Like MC methods, TD method is a simulation based model-free method that can learn directly from raw experience without a model of the environment's dynamics. Like dynamic programming, TD method used bootstrapping to update estimates based in part on other learned estimates. Q-learning proceeds by incrementally improving its evaluations of the \emph{Q-values} that incorporate the quality of particular actions at particular states. The evaluation of the action-value pair, or the Q-value, is done by learning the \emph{Q-function} that gives the expected utility of taking a given action in a given state and following the optimal policy thereafter.

\emph{Application of Q-learning to routing and CRNs}: It has been known for long that RL algorithms can be applied quite naturally to the routing problem in communication networks \cite{boyan1994packet}. The `Q-routing' algorithm, proposed by Boyan in 1994, learned a routing policy that minimizes total delivery time through experimentation with different routing policies. Some desirable features of this approach were: \emph{i)} its learning is continual and online, \emph{ii)} it is robust in the face of dynamic network conditions, and \emph{iii)} it is distributed and uses local information only. This early paper established that adaptive routing is a natural domain for reinforcement learning and a lot of follow-up work has taken place. Q-learning is perhaps the most popular model-free reinforcement learning technique which has been applied to CRNs extensively \cite{bkassinysurvey}. We refer the interested reader to the survey papers \cite{bkassinysurvey}\cite{al2013application} for more details and references.

\vspace{2mm}
\emph{Learning Automata (LA):} LA is a RL technique that subscribes to the policy iteration paradigm of RL which, unlike other RL techniques, operates by directly manipulating the policy $\pi$ \cite{nicopolitidis2011adaptive} \cite{akbari2010mobility} \cite{akbari2010intelligent}. A learning automaton is a finite state machine that attempts to learn the optimal action (that has the maximum probability to be rewarded) while interacting with a stochastic environment. The application of LA techniques for building adaptive protocols in CRNs is particularly appealing due to the natural simplicity of the LA approaches, and the general applicability of RL techniques to CRNs. Accordingly, LA has been used in the design of wireless MAC, routing and transport-layer protocols \cite{nicopolitidis2011adaptive}.

\emph{Application of LA to routing and CRNs}: Torkestani et al. have proposed using LA for multicast routing in mobile ad-hoc networks or MANETs\footnote{MANETs share an important characteristic with CRNs in that both of them have highly dynamic topology. The dynamically changing topology in MANETs is due to node mobility while in CRNs it is due PU arrivals.} to find routes with expected higher lifetimes through prediction of node mobility \cite{akbari2010mobility}. Another LA-based distributed broadcast solutions can be seen at \cite{akbari2010intelligent}.

\subsection{Game Theory}
\label{sec:gametheory}

Game theory is a mathematical framework which can be used to model interactive decision making between multiple decision making entities. Although, game theoretic models exist for both cooperative and non-cooperative settings, it is precisely the ability to model the \emph{competition between multiple agents} which distinguishes game theory from optimization theory and optimal control-theoretic frameworks such as the MDP \cite{haykin2005cognitive}. Game-theoretic analysis is relevant when the decision making is a result not only of environment, but also the decision of other `players' (or decision makers).

Multicast, particulary of multimedia content, incurs a significant cost to the network in terms of bandwidth and overall power consumption which has to be borne by the network nodes.  It is usually desired to distribute this cost across various receivers by devising some cost-sharing mechanism. Considering non-cooperative scenarios, it is possible for nodes to cheat and thereby maximize their personal utility. The framework of game-theory has presented itself as a viable choice for modeling the problem of selfish routing in CRNs \cite{mackenzie2006game} using \emph{``mechanism design''}, sometimes called reverse game theory, which allows us to devise appropriate mechanisms for a game such that rational players interested in maximizing their personal utility will play into a desired equilibrium point. Some of the works on cost-sharing mechanisms for multicasting in wireless networks are \cite{penna2004sharing} \cite{bilo2006sharing} \cite{singh2011wireless} \cite{panda2012wireless}.

\section{Protocols used for Wireless Multicasting}
\label{sec:multicastprotocolsinCRNs}

Multicast routing for wireless networks is an active area of research, and protocols addressing various issues such as energy efficiency \cite{maric2005cooperative}, throughput maximization \cite{zeng2007multicast}  \cite{zeng2010efficient} \cite{ramamurthi2009multicast}, and delay minimization \cite{hoang2009channel} \cite{isazadeh2010traffic} have been proposed in literature. Although our focus in this paper is multicasting in CRN, it is also prudent to review related work in multicasting which has focused on general multi-hop wireless networks since such networks share many common attributes with CRNs. We will discuss such protocols which have been proposed for wireless networks in general in section \ref{subsec:multwireless}. We will present protocols proposed specifically for CRNs in \ref{subsec:multCRN}.

\begin{center}
\begin{table*}
\small
\caption{\textbf{Representative summary of the various protocols and algorithms proposed for multicasting in CRNs}}
    \begin{tabular}{  m{3.9cm}  m{2.3cm}  m{4.3cm}  m{6.3cm}  }
  \toprule
    \textbf{Proposed work} & \textbf{Technique/ \mbox{Algorithm}} & \textbf{
		Multicast objective} & \textbf{Summary} \\
  \midrule
  \multicolumn{4}{l}{\textbf{Scheduling work}} \\
    AMS \cite{almasaeid2010assisted} & Scheduling & Min. total multicast time  &
	Proposes an assistant strategy to reduce the effect of channel heterogeneity and thereby improve multicast throughput performance.
   \\
  \multicolumn{4}{l}{\textbf{Network Coding based work}} \\
   \\
    Multicast scheduling with network coding \cite{jin2010multicast} &	Network \mbox{Coding} &	Improve multicast performance by \mbox{efficiently} utilizing channel resources & 
	Greedy and online protocol that provides QoS guarantees\\
   \\
  \multicolumn{4}{l}{\textbf{Optimization based work}} \\
   \\
	OMRA \cite{almasaeid2010demand} &	Dynamic \mbox{Programming} & Reduce the end to end delay and throughput degradation &	
    Dynamic programming based solution for optimal channel allocation.\\
		
	Channel allocation and multicast routing in CRNs [Shu et. al] \cite{shu2013channel}
	 & MILP & Max. multicast throughput &	
	 Joint channel assignment and multicast routing solution that models PU activity and interference      \\

		
	Scalable video multicast \cite{hu2010scalable} & MINLP &	Optimize the received video quality and ensure fairness among users &	
        Formulates video multicasting as an mixed-integer NLP problem. A sequential fixing algorithm, and greedy algorithms are proposed.
        \\

		
	Multicast communication in multi-hop CRNs [Gao et. al] \cite{gao2011multicast} & MILP &	Reduce network wide resources to support multicast sessions	&
        Formulates multicasting as a mixed-integer LP problem via a cross-layer approach and provided a polynomial-time centralized heuristic solution. \\

 \\ 
 \multicolumn{4}{l}{\textbf{Tree Construction Techniques}} \\
	COCAST \cite{kim2009cocast} & Source Based Tree &	Improve the scalability of ODMRP &	
        Reactive routing protocol incorporating channel assignment that seeks to maximize delivery ratio and minimize delay is proposed \\

		
	MEMT \cite{ren2009minimum} &	Steiner Tree  &	Construct minimum energy multicast tree &	
		 Proposed a low-complexity approximate solution
         \\

		
	QoS multicast \cite{xie2012qos} & Spanning Tree &	Tree construction with minimum bandwidth consumption &	
			A QoS-satisfying multicast tree, with minimal bandwidth consumption, is constructed through a novel slot assignment algorithm.
         \\
	\bottomrule
    \end{tabular}
		\label{tab:multicasttable}
		\end{table*}
\end{center}

\subsection{Multicast routing in wireless networks}
\label{subsec:multwireless}

A study in \cite{matam2013improved} solves the problem of multicasting in wireless networks with the objective of bandwidth conservation. A protocol has been proposed which computes the multicast trees with minimum bandwidth consumption through a heuristic approach. The performance evaluation shows that even in the worst case scenarios, the proposed algorithm works better compared to other baseline algorithms. A new degree of freedom has been explored in \cite{oh2013mr2_odmrp} by using multiple transmission rates with multiple radios. A new multicast routing protocol inspired from ODMRP is proposed to tackle the problems introduced by multi radios and multi rates transmission. Optimization technique is used and the problem is modeled as integer linear program to obtain the values for rates at every node and to construct the optimal tree. The results show that the proposed protocol produces solutions which are near optimum and also outperforms ODMRP in terms of end to end delay.

In \cite{prasad2013multiobjective}, An ant colony optimization approach is proposed to tackle the minimum cost multicast tree with delay and bandwidth constraints problem. A niched ant colony optimization with colony guides (NACOg) algorithm is proposed to tackle the problem. The results show the better performance of the NACOg as compared to other algorithms like Haghighat genetic and KPP heuristic in terms of finding the minimum cost QoS multicast tree.  In another research \cite{alasaad2013ring},  Qos multicasting performance in wireless mesh networks is studied  over a ring routing topology. An algorithm is proposed for IP multicasting routing with the mesh routers supporting the group communication. The proposed algorithm outperforms the traditional schemes in terms of end to end delay and capacity of multicast networks.

Learning automata has also been used to formulate the problem of multicast routing in wireless networks. In \cite{jahanshahi2013lamr}, the problem of channel assignment and multicast routing in wireless mesh networks is solved jointly in the learning automata framework. The performance evaluation depicts that the proposed scheme LAMR (Learning Automata based Multicast Routing) outperforms the well known algorithms LCA (Level Channel Assignment) \cite{zeng2007multicast} and MCM (Multi-Channel Multicast) \cite{zeng2010efficient} in terms of packet delivery ratio, delay and throughput.  In another study \cite{cheng2011joint}, QoS multicast problem merged with the channel assignment problem in wireless networks is presented. The authors claim that the multicast tree construction and channel assignment problem have been solved separately. They propose three different algorithms for finding minimum interference multicast tree with efficient use of scarce network resources.

\subsection{Multicast routing protocols in CRNs}
\label{subsec:multCRN}

CRNs are very dynamic in nature. Topology changes in CRN depend upon the location and activity of the primary users. These frequent changes cause a number of problems for multicasting in cognitive radio networks compared to multicasting in traditional wireless networks. Channel heterogeneity is one of such problem, where two neighboring nodes may not have a common channel available and they have to use different channels for multicasting. Thus as a result,a single multicast transmission is broken into many small unicast transmissions introducing significant switching delay. Achieving route stability is another problem, as CR transmissions need to be interrupted whenever a PU activity is detected. Routing algorithm should be capable of dealing with such changes and adopt accordingly. We will now be presenting the solutions that have been proposed to address the challenges that are faced while deploying multicast in CRNs. We will discuss these solutions, according to their applications or source technique, in categories of scheduling, network coding, optimization and efficient tree construction. A representative summary is also presented in table \ref{tab:multicasttable}.

To cope with the channel heterogeneity problem, \emph{multicast scheduling protocols} have been proposed.In \cite{almasaeid2010assisted}, assisted multicast scheduling is proposed to reduce the end to end delay. The main purpose is to minimize the total multicast time in a single cell of cognitive mesh networks. The proposed scheme uses three operations of assisting, overhearing and codeword exchange. The assistance operation allows the multicast receivers to assist in the process of multicasting and to forward the data to the other receivers too. Overhearing introduces the assistance between two different multicast groups. This happens when some nodes belonging to a group overhears a transmission intended for another group. After overhearing, these nodes can now forward the data intended for the other group. Codeword exchange is also introduced to assist the multicast scheduling by using coded packets. The intended multicast receivers can decode and extract their data easily. These three operations help in reduction of total multicast time. The results show the better performance of assisted multicasting in terms of throughput and total multicast time. It is shown that when no assistance is used, the sender uses six slots to transmit data intended for the multicast receivers. When intra group assistance is allowed through assistance operation, the number of time slots required to transmit data reduces to five. Furthermore, when the overhearing operation is enabled and inter group assistance is allowed, the number of slots reduce to four. Finally, the number of slots used to transmit data reduce to three when \emph{network coding} is used and coded packets are exchanged. In another study \cite{jin2010multicast}, a greedy scheduling protocol is proposed to optimize the overall performance of the network. In this work, fairness among users is also considered along with the efficient utilization of spectrum. A cooperative transmission link is allocated with at most one channel so that more cognitive users are encouraged to participate in the cooperative communication. They have also adopted the network coding so that the overhead can be reduced and better error control can be performed. The problem is formulated as non linear integer program and an online scheduling protocol is introduced for channel allocation and power control policies. The proposed protocol allocates channels to the links according to a distributed algorithm which is claimed to be a good operator in realistic systems.

Multicasting problems have also been studied using methods of \emph{optimization theory} (which was introduced in section \ref{sec:optimization}) in a few studies. The \emph{dynamic programming} approach has been used for the problem of multicasting in CRNs \cite{almasaeid2010demand}. The authors have considered the problems of broadcast deformation and channel switching delay while proposing a multicast algorithm for cognitive radio networks. The problem of broadcast deformation is very common in cognitive radio networks because of their dynamic environment due to PUs. The set of available channels to a cognitive user depends upon the channel occupancy of the primary user. Therefore, the available channels set for every user might be different which can deform a broadcast transmission into a few multicast transmissions or many unicast transmissions in the worst case scenario. The switching delay is also introduced due to the different channels at different users which causes a user to continuously switch between channels and hence channel switching latency is introduced. This study proposes the algorithms to tackle the problem of channel heterogeneity and latency introduced due to the channel switching. The simulations show the better performance of proposed algorithms in terms of delay as compared to baseline algorithms. In \cite{shu2013channel}, the multicast routing and channel allocation is jointly formulated as an optimization problem. The problem is carried out with the objective of increasing throughput per session. The results throw light upon the channel selection in cognitive radio networks.

\emph{Cross layer optimization} approach has been used for the problem of multicasting in cognitive radio networks in \cite{hu2010scalable} and \cite{gao2011multicast}. In \cite{hu2010scalable}, cross layer optimization approach is used to multicast data to the multimedia receivers while optimizing the received video quality. The proportional fairness among multicast receivers and interference minimization are also taken into account. A sequential fixing algorithm and a greedy algorithm is proposed in order to achieve the fairness and to minimize the interference while optimizing the video quality. Linear relaxation of the mixed integer non linear programming problem is used by the fixing algorithm and the greedy algorithm makes use of the inherent priority structure of fine granularity structure video and ordering of user channels according to their qualities. To adjust the calculated solution according to updated channel sensing results, a less complex greedy algorithm is used. The results show that the proposed algorithms performs better than the baseline algorithms in terms of average peak to signal to noise ratio. In \cite{gao2011multicast}, cross layer optimization approach is used with the purpose of reduction in the required network resources. A self interference constraint protocol is proposed which assigns a frequency channel at each transmitter to at most one session of multicast. These frequency bands are assigned by using the self interference constraint in which the receiving nodes are not allowed to receive simultaneously from two transmitters. In this way, interference is minimized. The purposed algorithm solutions are compared with the lower bound which show that the proposed solution provides near optimum values.

Some \emph{CRN-specific multicast routing protocols} have recently been proposed. In \cite{kim2009cocast}, the authors have proposed a multicast routing protocol for CR-equipped MANETs to alleviate the scalability issues associated with the well known multicasting protocol ODMRP \cite{lee2002demand}. All the nodes are considered mobile with a single radio per node and multiple channels are available in the network. COCAST works in similar fashion as ODMRP works but the difference lies in construction of tree and mesh. The ODMRP constructs a mesh so that alternative paths can be used upon the breakage of routes. In contrast to ODMRP, COCAST constructs a tree which is different from mesh in terms of alternative paths and reduces the route overhead. The simulation results show that the proposed algorithm COCAST performs better when number of multicast sources increase in the network. The proposed algorithm is more scalable as compared to the ODMRP.

The problem of multicasting has also been studied as a \emph{minimum energy multicast tree construction} problem \cite{ren2009minimum}. The purpose is to save the overall energy used in the construction of a multicast tree. Apart from the energy consumed in tree construction, energy is also consumed in spectrum sensing and transmission of data. These energy consumptions are also taken into the account. The multicast problem is formulated as directed Steiner tree problem in this study. An approximation algorithm is proposed for this purpose and results are computed to see the impact of primary traffic load on the minimum energy multicast tree. The results show the better performance of this algorithm as compared to the approximation algorithm \cite{liang2006approximate} proposed for the traditional networks. Apart from the minimum energy, a protocol has also been proposed for construction of a tree with minimum multicast bandwidth consumption \cite{xie2012qos}. Two methods have been proposed in this study for minimum bandwidth tree construction. The first methods constructs the minimal spanning tree first and then slot assignment is done through a proposed algorithm. where as the second approach considers these two problems of minimal spanning tree construction and slot assignment jointly in such a way that overall bandwidth consumption is minimized. The metrics used for the evaluation of proposed algorithm are transmission of slots and success rate and these metrics show the better performance of the algorithm over the baseline algorithms.

\section{Open Research Issues and Future Directions}
\label{sec:openissues}

\subsection{Incorporation of AI techniques into Multicasting Framework}

Since CRNs often have to operate in dynamic unpredictable and unknown environments, it is important to integrate AI-based techniques seamless into the core of the routing framework. The design of \emph{``cognitive multicasting protocols''} that can intelligently adapt to changing network conditions is an important, and yet unexplored, area of research. The interested reader is referred to a detailed tutorial and survey on the topic of AI-based cognitive routing protocols for CRNs \cite{qadir}.

\subsection{Building Reliable Multicast Routing Protocols}

The ARQ scheme adopted in unicast reliable end-to-end protocol is not suited to multicast since it requires every packet, or a group of packets, to be ACK'ed by the receiver. Using a negative acknowledgment mechanism (NACK) with a semantic of a retransmission request is better suited for multicast transmission \cite{diot1997multipoint}.

\subsection{Incorporation of Spectrum Modeling}

The multicast routing framework should incorporate spectrum modeling into its basic design. One way is to probabilistically model the PU arrival process and traffic pattern to avoid the channels that will be claimed by PU with a high probability. For example, a SU can exploit spectrum sensing data to select white spaces (that emerge due to absence of PUs) that tend to be longer lived at a particular time of the day and a particular location.  A number of techniques have been proposed for spectrum prediction including techniques that are: \emph{i)} Hidden Markov model based, \emph{ii)} Neural Networks based, \emph{iii)} Bayesian inference based \cite{xing2013spectrum}. For more details about spectrum prediction techniques, the interested readers are referred to a detailed survey on this topic \cite{xing2013spectrum} and the references therein.

\subsection{Incorporation of Extra Degrees of Freedom}

A future research direction is incorporating extra degrees of freedom such as mobility, link-layer rate diversity, interface diversity seamlessly with techniques such as network coding, game theory, and optimization.

\subsection{Conclusions}
\label{sec:conclusions}

There has a lot of work on wireless multicasting. In particular, various \emph{algorithms} for developing multicast forwarding structures have been proposed, and various \emph{techniques}, including the frameworks of optimization theory, network coding, heuristic techniques, etc. have been used to improve multicast routing. Multicast \emph{protocols} incorporate both techniques and algorithms to define the rules, syntax, and semantics of multicast communication. In this paper, we have presented a coherent account of the overall landscape of multicasting algorithms, techniques, and protocols that apply to multi-hop cognitive radio networks. In particular, we have presented both a self-contained tutorial of multicasting algorithms, techniques, and protocols that apply to CRNs along with and a detailed survey of their applications. We have also identified open research issues, and have identified promising directions for future work.

\bibliographystyle{ieeetr}
\bibliography{ReferencesFile}

\epsfysize=3.2cm

\end{document}